# Collaborative design:
# Managing task interdependencies and multiple perspectives

Françoise DÉTIENNE
*EIFFEL Research Group*
*INRIA - Domaine de Voluceau, Rocquencourt*
*B.P 105, 78 153 Le Chesnay Cedex, France*
e-mail: *Francoise.Detienne@inria.fr*

**Abstract**

This paper focuses on two characteristics of collaborative design with respect to cooperative work: the importance of work interdependencies linked to the nature of design problems; and the fundamental function of design cooperative work arrangement which is the confrontation and combination of perspectives. These two intrinsic characteristics of the design work stress specific cooperative processes: coordination processes in order to manage task interdependencies, establishment of common ground and negotiation mechanisms in order to manage the integration of multiple perspectives in design.

## 1. Introduction

Studies on reasoning in design have usually been carried out on individual problem solving activities. In response to the increasing need to assist collective work in an industrial context, more recent studies have shifted their foci toward cooperative work. A major concern in industrial modernisation is the creation of new technico-organisational systems which support collective work, greater interaction between design stakeholders, as well as capitalisation and reuse of design knowledge.

The switch of focus in design studies, from individual to collaborative design, has entailed the emergence of research issues, with respect to cooperative work. These research issues have been developed with respect to the challenges and particularities in the design application domains addressed in the design studies. In this paper we will adopt a generic approach on design focused on mechanisms which we assume independent of the application domain.

This paper focuses on two characteristics of collaborative design with respect to cooperative work: the work interdependencies, linked to the nature of design problems, which are most important in distributed design; and the confrontation and combination of perspectives which are most important in co-design.

In the following of this paper, we will first outline a framework for collaborative design. Then we will develop cooperation issues related to our two foci: collaborative design as managing task interdependencies (modular design and work organisation, informal communication and roles, awareness) and collaborative design as managing multiple perspectives (establishment of common ground, perspectives clarification and convergence mechanisms). In this paper we will discuss approaches for supporting collaborative design which are not only technical but, most often, technico-organisational.

## 2. A framework for collaborative design

### 2.1. *Design as an activity*

Design is an activity consisting in specifying an artefact, given requirements that indicate one or more functions to be fulfilled and/or objectives to be satisfied by the artefact (Visser, 2004). The most common conception of design problems is to consider them as "ill-structured" problems (Eastman, 1969; Simon, 1973). Their characteristics are as follows.

There are many degrees of freedom in the problem's initial state. The requirements given at the start are never complete or unambiguous: initial problem specifications are not sufficient to define the goal, i.e., the solution, and progressive definition of new constraints is necessary.

Design requirement, in terms of constraints and subgoals, are unstable. Instability of the design requirements, are due to external causes, e.g., changes introduced during design by the client. Instability of subgoals are caused by opportunistic planning which leads to creation and modification of subgoals in design (Hayes-Roth and Hayes-Roth, 1979). This way, both the problem representation and the solution representation continuously evolve during design.

A design problem has several acceptable solutions. There is no definite criterion for testing any proposed solution to a design problem, such as there typically exists for "well-structured" problems: various solutions are acceptable, one being more satisfactory according to one criterion, another according to a different criterion; that is, design problem solutions are not either "correct" or "incorrect, they are more or less "acceptable".

Design problems tend to be large and complex. They are not generally confined to local problems, and the variables and their interrelations are too numerous to be divided into independent sub-systems. In individual design, the division of problems into sub-problems is supposed to result in a reduction of complexity —often, however, design problems are difficult to decompose into independent sub-problems, and different decompositions of a same problem are possible. In collaborative design, which can be viewed as a paradigmatic case of tightly coupled work (Olson and Olson, 2000), this complexity produces great work interdependencies. One consequence of this complexity is also that solving these problems often requires that multiple competencies be put together, which in turn leads to development of collaboration between co-designers from various disciplines and thus involves the management of multiple perspectives.

The switch of focus in design studies, from individual to collaborative design, has entailed an evolution of the theoretical frameworks. The purely cognitive framework based on the information processing model is not sufficient to address issues related to the collective nature of work. Bucciarelli (1988) considers that designing is more than a cognitive process, although design knowledge and designers' heuristics are essential ingredients. Both Bucciarelli (1988) and Schön (1988) advance that designing is a social process. Research theoretical frameworks in design studies have integrated the social and organisational aspects and the situated aspects of the collaborative design situations. This has led to more socio-cognitive frameworks, with the adoption of concepts developed in distributed cognition, situated cognition, and activity theory.

*2.2 Design domains*

Design covers a great range of activities in various knowledge domains: engineering design, architectural design, software design, etc. Historically, various research issues have been focused on depending on the particularities of design domains.
In software design and engineering design studies, in which design is guided (if not constrained) by process models and methods, the issues have covered work interdependencies, modular design and coordination issues. In architectural design and other domains like mechanical design in which there are various forms of representation of the design artefact, in particular graphical ones, one important issue has been to understand the roles of these representations in design.  In design of products, where taking into account users or end-users is an important aspect of design, one important issue has been to anticipate the uses by collaborative methods such as participatory design.
As stressed by Thomas and Carroll (1979) design activities involved in these various domains have much in common. This is the approach we follow in this paper, assuming that there are similar cooperative processes whatever the application domain. However, some design domains, in which some processes are emphasized, may be heuristically more relevant to study these processes.

*2.3 Design and HCI*

Two relationships between design and HCI may be stressed out. Firstly, the designer can be considered as an end user if we consider the design activity itself and the technical support to it. It is the approach we have adopted in this paper. We have considered the design process going from the initial specifications to the production[1] and maintenance of the design artefact and the tools to support design collaboration during this process.
Secondly, we can consider the end-user of the designed artefact. End-users may be more or less involved in the design process depending on the design approach. In classical user-centred design situations, the role of the user is informative (e.g., needs elicitation) and evaluative (e.g., prototype evaluation), whereas in participatory design situations, the role of the user is also generative (solution elaboration) and sometime decisional: this way, the end-user becomes a co-designer.

---

[1] We usually consider that design activity ends when the constructed representation of the artefact is precise and detailed enough that the implementation can take place. However, the implementation phase (production, maintenance) can also involve decisions which will change the representation of the artefact specifications.

This approach takes into account the continuation of design in usage. The users appropriation of artefacts as instruments has been covered in a recent special issue of IWC (vol 15). Based on the activity theory, human instruments are seen as containing components from artefacts themselves and components from users' utilization schemes (Rabardel and Bourmaud, 2003). The user, through its use, turns the artefact into an instrument. Extending this approach, design can be viewed as a mutual learning process between users and designers (Béguin, 2003). We have left aside this (important) current field of research in our paper.

*2.4 Distributed design versus co-design*

Falzon (1994) has stressed a distinction between two design situations according to the nature of shared goals: co-design and distributed design. A design project can be characterised by cycles of distributed design and co-design. Each design situation has specific needs of assistance and specific relationship with cooperative processes.

In distributed design (a notion different from distributed work and distributed cooperation), each actor or team has his/her own task to perform. They pursue goals (or at least sub-goals) that are specific to them. They do more than share resources. They need to coordinate in order to pursue their own tasks because of interdependences between tasks of the design project. Managing work interdependencies is one great challenge in computer-supported cooperative work. Components of the design work are highly interdependent thus leading to coordination needs in collaborative design.

In co-design, actors or teams share an identical goal and contribute in order to reach it through by applying their multiple perspectives. They do so with very strong constraints of direct cooperation so as to guarantee a solution to the problem. Design is a process of negotiating among disciplines. Solutions are not only based on purely technical problem-solving criteria. They also result from compromises between designers: solutions are negotiated (Bucciarelli, 1988). In terms of the cooperative work arrangements distinguished by Schmidt[2] (1994a), design typically serves a generic function which is the confrontation and combination of perspectives. It is a cooperative work arrangement which facilitates the application of multiple perspectives (conceptualisation of the field of work) on a given problem so as to match the multifarious nature of the field of work.

Thus, two characteristics of collaborative design are more or less central depending on the design situation: whereas managing task interdependencies is most central in distributed design, managing multiple perspectives is most central in co-design. These two intrinsic characteristics of the design work stress specific cooperative processes: coordination processes in order to manage task interdependencies, establishment of common ground and negotiation mechanisms in order to manage multiple perspectives in design.

---

[2] The other work arrangements identified by Schmidt are:
- Augmentation of capacity: a cooperative work arrangement which augments the mechanical and information processing capacities of human individuals. For example, moving a piano.
- Differentiation and combination of specialties: a cooperative work arrangement which combines multiple technique-based specialties. For example, a surgery operation.
- Mutual critical assessment: a cooperative work arrangement which facilitate the application of multiple problem-solving strategies and heuristics to a given problem and ensure balanced and objective decisions. For example, medical diagnosis.

The spatio-temporal (ST) framework (Johansen, 1988) has often been taken in the literature to differentiate collaborative situations and tools supporting them[3]. Indeed, design as other collaborative work activities is more and more engaged in distributed frameworks where teams work in different spatial framework (co-location and distant) and in different temporal framework (synchronous and asynchronous). Design is increasingly a multi-site, multi-cultural, globally distributed undertaking. Design projects tend to become more and more geographically distributed: relationships between various distant teams and with subcontractors are typical of this situation. Multi-site (global) development tasks take much longer than comparable collocated tasks and collaboration plays a major role in this delay (Herbsleb and Mockus, 2003). We will consider that there is no systematic relationship between cooperative processes and the ST framework. The ST framework will not determine what cooperative processes will occur but will rather influence the way it occurs and in which extent it occurs.

### 3. Collaborative design as managing task interdependencies

Managing task interdependencies is an important issue, well identified in Computer-supported cooperative work (Schmidt, 1994b): actors engaged in cooperative work are mutually dependent in their work. "Being mutually dependent means that A relies positively on the quality and timeliness of B's work and *vice versa*". This has to be opposed to "distributed cooperation" where operators share an information space but are not necessarily conscious of other operators work.

In collaborative design, it is in the distributed design phases where each designer or team works on specific subtasks, that managing task interdependencies becomes more crucial. It is also emphasised by work at distance as compared to co-location (such as in open-plan offices). It involves aligning work via various co-ordination mechanisms.

Several socio-technical solutions are possible, in particular:
- At the task and organisational level: matching organisation structure with task decomposition.
- At the cooperation level: facilitating coordination mechanisms through informal communication, and awareness.

*3.1. Coupling of work, modular design and work organisation*

In project design, tasks corresponding to sub-problems are distributed among individuals or teams, each carrying out various sub-tasks with great interdependencies between these subtasks. This is related to the concept of "coupling of work" (Olson and Olson, 2000). Typically, most collaborative design tasks are tightly coupled. Olson and

---

[3] In general, various work modes, synchronous or asynchronous and collocated or distant, alternate in design projects. Purely asynchronous distant work may be found in Open Source Software (OSS) design projects in which co-designers coordinate their activity almost exclusively by means of email and bulletin boards (Sack et al. 2005; Ducheneaut, 2003; Gasser et al. 2003; Mockus, Fielding & Herbsley, 2002).

Olson (2000, p 163) even mention that: "Design can be a paradigmatic case of tightly coupled work" even if some design process can make it more routine.

Coupling refers to the extent and kind of communication required by the work. Olson and Olson (2000, p 162) define this concept in relation with the concept of decomposability of systems in organisational theory: "Tightly coupled work is work that strongly depends on the talents of collection of workers and is non routine, even ambiguous. Components of the work are highly interdependent. The work typically requires frequent, complex communication among the group members, with short feedback loops and multiple streams of interaction."

One socio-technical solution to managing task interdependencies has been to reduce it and to match as far as possible the task decomposition with the organisation structure.

In Software design, particular design methodologies such as modular design (Parnas, 1972) aim to reduce interdependency, thus transforming tightly coupled work into loosely coupled work as far as possible. A module is a responsibility assignment rather than a sub-program which implies that dividing a software system is simultaneously a division of labour. According to Conway's law, the structure of the system mirrors the structure of the organisation that designed it. Conway's law (Herbsleb and Grinter, 1999) was the first explicit recognition that the communication patterns left an indelible mark upon the product built. Following this approach, Olson and Olson (2000) recommend to design the work organisation so that ambiguous, tightly coupled work is collocated. This should keep the need for cross-site communication at a minimum.

One limit of this approach is that Parnas and Conway focused only on the structure of the product, which indeed provides an important foundation for the coordination of work. However there is also dependency with respect to the process. Two other types of interdependency could be distinguished in the design process:

- Between refinement levels (aggregation hierarchy): it is well identified by Herbsleb and Grinter (1999): e.g. between different stages of design: conceptual design, coding, testing. For example, process balanced development breakdowns may appear because designers from different disciplines attach different levels of priority to particular sub-problems (Martin et al. 2002a) which entail a gap between the refinement level of the design artefact, which is processed by the various disciplines at a particular time.
- Between abstraction levels (abstraction hierarchy): functional, structural, operational (Rasmussen, 1979).

Another socio-technical solution for managing task interdependencies has been to develop workflow or project management tools in order to guide and monitor progress through a task: task goals, task decomposition into subtasks, dependencies among tasks and subtasks, actors roles and assigned responsibilities, tasks and subtask completion status. These tools are aimed to support coordination. They have at least two kinds of limit for design task:

- Design work is a tightly coupled work with high and complex dependencies between tasks. Modelling these dependencies in a workflow system is not easy and even not possible.
- Design work involves opportunistic planning and re-planning (Hayes-Roth and Hayes-Roth, 1979) : opportunistic planning leads to creation and modification of goals and subgoals. This evolution is not taken into account in workflow systems in which the task planning is based on the prescribed process model rather than on the effective activity.

*3.2. Informal communication and informal roles*

Formal communication, e.g., formal meetings, and informal communication, e.g. unplanned meetings or discussions, are very important to ensure coordination. Communication on the state of design specifications and solutions, design decisions, design plans produced by various designers may be delayed or not happen. In an earlier study of software design, Krasner et al (1987) distinguished various communication breakdowns:
- No communication between groups that should be communicating;
- Miscommunication between groups;
- Groups receiving conflicting information from multiple sources;
- Communication problems due to design dynamics (changes of people, goals, technology).

Previous studies stress the importance of informal communication to ensure coordination. Informal communication takes place as needed. It can be ensured by people playing informal roles in the organisation: boundary spanners (Grinter, 1999; Krasner et al. 1987) also referred to as "contact people" or "liaison" (Herbsleb et Grinter, 1999). Boundary spanners are people who move among different teams transferring information about the state of the project. They translate information from a form in which it was used by one team into a form that could be understood by other teams. They reduce the amount of information lost or miscommunicated between different phases of design development and different development teams. Boundary spanners are characterised as an informal role, adopted by persons with good communication skills who have contacts with various teams. Krasner et al (1987) observed that, when individual appointed to formal liaisons positions did not have skills that matched the needs of boundary spanners role, a different person usually emerged to fill it informally.

Depending on the organisation and the spatio-temporal framework, informal communication may become quite difficult and even impossible. Collocation, as in open-plan working areas, encourages informal communication. In distant design, several barriers to communication, leading to coordination breakdowns, may be outlined (Herbsleb and Grinter, 1999; Herbsleb and Mockus, 2003; Krasner et al. 1987; Martin et al. 2000a; Olson and Olson, 2000).

One barrier refers to the lack of willingness to communicate. Lack of trust and willingness to communicate openly is linked to collaboration readiness of the organisation, and, in multi-site situations, to the perception people have to be part of the same team. Also, communications are often poorly rewarded whereas there may be strong internal rewards for remaining an "expert" by possessing knowledge that others on a project do not have and by dispersing it slowly.

Another barrier refers to the difficulty to establish contact. Firstly, "knowing who to contact about what" may be problematic, in particular, in multi-site design: people do not know each other personally and cannot identify easily who is the person who has the relevant information or is the expert. Some tools for finding experts, such as the Expertise Browser (Mockus and Herbsleb, 2002) could provide a technical solution.

Secondly, the cost of initiating contact may be too high, in particular, with respect to availability and responsiveness of people. Technical solutions could be tools to support contact, like NetMeetings and to support presence or social (or peripheral) awareness, like media spaces. Furthermore, technology readiness (Olson and Olson, 2000) will favour the adoption of such tools in an organisation.

Another barrier refers to the difficulty to communicate. Firstly, designers may not have communication skills. As seen above this kind of ability is a key characteristic of boundary spanners. Also the culture of sharing and collaborating in an organisation (referred to as collaboration readiness by Olson and Olson) is a key characteristic in the adoption of tools to communicate. Secondly, designers may not have shared representational formats on which to communicate. Many different representational methods may be used to organize and communicate a shared model of system. More and more, design projects tend to establish a common representational format to facilitate communication. Thirdly, the lack of local jargon should make communication harder.

*3.3. Awareness*

Awareness is a concept which has been developed mostly in the process control paradigm, in particular in studies on coordinative practices in shared setting such as control rooms (see for example, Heath and Luff, 1992). More recently, this concept has started to become of interest in the design paradigm, e.g., awareness in open-plan working areas referred to as warrooms in (Teasley et al., 2000), and also in geographically distributed contexts. Displaying selectively information for distributed groups via shared electronic and physical media may support awareness in remote collaboration (Everitt et al. 2003). This becomes an important topic in design platforms.

Awareness denotes those practices through which actors tacitly and seamlessly align and integrate their distributed and yet interdependent activities (Schmidt, 2002). It refers to practices through which cooperating actors, while engaged in their respective individual activities and dealing with their own local urgencies and troubles, manage to pick up what their colleagues are doing or not doing and to adjust their own individual activities accordingly. This definition fits well to what we refer to as the distributed design situation (as opposed to co-design) in which co-designers fulfil their own tasks while conscious that a more general design task, for which they need to coordinate, is shared with co-designers.

Awareness support must include facilities for monitoring, directing attention, and handling over tasks at different levels of obstrusiveness and persistence (Carstensen and Schmidt, 1999). It involves two selective processes: displaying and monitoring which are complementary aspects of coordinative practices (Schmidt, 2000). Actors display "selectively" what they do, e.g., display what they assume relevant to colleagues in the particular situation, in a form and a level of granularity which is adapted to the situation facing their colleagues. They make their activities « publicly visible », available and accessible to others. This way, they put some information into the field of attention of others. Similarly, actors do not aimlessly monitor what happens around them. Actors scan for certain cues or indicators of states or states changes important for what they are doing or will be doing or could be doing.

Awareness is only meaningful if it refers to a person's awareness of something. Depending on the type of referent, Carroll et al. (2003) distinguish several awareness mechanisms for remote collaboration: social awareness, action awareness and situation awareness.

Social awareness (who is around?) is a weak conception of awareness with respect to cooperation. It is awareness of the social context of work, not on the on going activities and artefacts of a joint cooperative effort (Schmidt, 2002). However, as noted in section 3.2, social awareness, e.g. detecting some actor presence and availability (through, for example, mediaspaces), can engender informal communication which may serve design coordination.

Action awareness (what is happening?) refers to the state of task-oriented objects and collaborators contributions: timing, type or frequency of collaborators' interaction with a shared resource; location and focus of collaborators' current activity. It is oriented toward product dependency rather than process dependency.

Designers share information about the design project in progress through data repositories and knowledge management tools. Displaying the results of one's own design task involves updating regularly these information spaces. This becomes most important for coordination purposes. Designers may not update information spaces regularly because of lack of collaboration readiness, training, cost of information capture and lack of organisational support. One obstacle can be the necessity to shift from one application (usual to the designer) to another (not familiar) in order to entry data in the shared space without immediate feedback for this extra-task (Martin et al. 2000a). Monitoring the changes relevant to one's own work may also be too costly. Tools supporting action awareness are notification systems and systems for visualising what has been changed in shared spaces.

One important evolution of these tools would be to be "actor-sensitive" or "context-sensitive", i.e. displaying relevant information according to an actor's present task and goal. As noted by Cadiz et al. (2000) these systems should inform without overwhelming and separate higher and lower priority information for different actors at different times. This notion of actor remains also a research question: it is generally matched to the notion of functional role which is static in the project whereas roles also been defined as a construction made through the interactions.

Situation awareness (how are things going?) is awareness of other's people's plans and understandings: shared plans; assignments or modifications of project roles; tasks dependencies based on roles, timing, resources, status of design project progress. It is oriented toward process dependency.

Understanding and supporting situation awareness remains a great challenge for tightly coupled collaborative tasks such as design, in particular in remote settings. It is linked to the concept of workflow or project management. However, these tools are based on prescribed design models which do not take into account the evolution of the design model in practice, in particular, through opportunistic planning. Indeed, situation awareness depends on knowledge of what one's collaborators are effectively doing, rather than what a symbolic model, the "prescribed" design model, says they should be doing.

Understanding the designers practices, in design platforms, could be a way to progress on this topic: for example, in OSS platforms, understanding the practices of cross posting and the strategies of information search of highly active and competent contributors. As noted by Schmidt (2002), awareness is not the product of passively acquired information but a characterisation of some highly active and highly skilled practices.

## 4. Collaborative design as managing multiple perspectives

Design projects involve designers from various disciplines. In aeronautical design projects, participants are design office disciplines (structure, system installation, stress), downstream designers (maintainability, production) and also new disciplines that have appeared with the introduction of CAD (Computer Aided Design) and PDM (Product Data Management) tools.  In software design, there are "specialist groups": technical programming teams, configuration management, quality assurance, technical writing, contract management, customer representatives, systems engineering, hardware engineering, integration/test.

According to Schmidt's typology (1994a), design is a cooperative work arrangement which facilitates the application of multiple perspectives on a given problem. Design is the business of a collective or team whose different participants, with different competencies, responsibilities and interests see the object of design differently (Bucciarelli, 2002). They inhabit different "object-worlds". Design is a process of negotiating among disciplines. The design that results from such a process is a "social construction" (Bucciarelli, 1988). In co-design, reaching an agreement on solutions is not only based on purely technical problem-solving criteria. It also results from compromises between designers: solutions are negotiated.

This approach stresses the importance of multi-disciplinarity or multi-expertise in design and its implication. In co-design, two cooperative processes are of major importance:
- Establishment of common ground also called cognitive synchronization or construction of a frame of reference which ensures inter-comprehension. Here we will limit grounding to explanation (or clarification) mechanisms;
- Perspectives clarification and convergence mechanisms. Here, we will refer to argumentation mechanisms.

*4.1. Establishment of common ground*

Common ground refers to that knowledge people (who interact each other) have in common and they are aware that they have it in common. The establishment of common ground ("grounding") is an important process in design tasks because of the domain and cultural differences of co-designers.

Many studies have stressed out the relative importance of grounding in co-design meetings. We will make a brief overview of these studies in which grounding has been analysed through conversations taking place in collocated design situations, seen as a reference situation for this process, in particular, because of multi-modality and shared context. Then we will discuss of the establishment of a shared context and of grounding with respect to distant design.

*4.1.1. Importance of grounding*
Studies on face-to-face co-design meetings - in the development of software (Herbsleb et al. 1995, Olson et al. 1992, D'Astous et al. 2001), of aerospace structures (Visser, 1993), of mechanical devices (Stempfle and Badke-Schaub, 2002) or of a backpack-to-mountain-bike attachment (Cross et al. 1996) - either aimed at elaboration/brainstorming or evaluation (e.g. inspection meetings in software development) tend to show the predominance, in such meetings, of establishment of common ground.

This activity ensures inter-comprehension and construction of a shared representation of the current state of the problem, solutions, plans, design rules and more general design knowledge. The establishment of common ground is a collaborative process (Clark and Brennan, 1991) in which the co-designers mutually establish what they know so that design activities can proceed. Grounding is linked to sharing of information through the representation of the environment and the artefact, the dialog, and the supposed "pre-existing" shared knowledge.

In collocated design meetings, other types of activity, related either to the object of the design task or to the design process, occur. They concern:
- the design problem and solution:
    - design activities, i.e. elaboration, enhancement of solutions and of alternative solutions;
    - evaluation activities, i.e. evaluation of solutions or alternative solutions, on the basis of criteria.
- the group management:
    - project management activities, i.e. allocation and planning of tasks;
    - meeting management activities, i.e. ordering, postponing of topics in the meeting;

Olson et al. (1992) analysed the interactions of a group of experienced software designers during design meetings. They found that a great amount of time was being spent on design discussions involving the design themes and their clarification. In a study on software technical review meetings, D'Astous et al. (2004) emphasize the importance of cognitive-synchronisation exchanges (one third of the exchanges) as compared to review exchanges, which are supposed to be the main objective of these meetings. They show that construction of a shared representation of the to-be-reviewed design solution is a prerequisite for evaluation activities to occur: evaluation is indeed often introduced by cognitive synchronisation. Stempfle and Badke-Schaub (2002) found that some teams bypassed grounding (referred to as "analysis") and that this led them to premature evaluation of design ideas. However, the fact that the observed teams were composed of students in mechanical engineering may partly explain this bias toward premature evaluation.

Grounding concerns not only the problem and solution, but also the review procedure and the representational conventions. D'Astous et al (2004) observed that co-designers make explicit their criteria and order them, which reveals construction of common knowledge, required in context, on the evaluation procedure. Curtis et al. (1988) observed that the process of a team coming to common representational conventions could take, in early phases of development, as much time as did the use of the conventions themselves.

*4.1.2. Shared context, distance and asynchronicity*
Collocation is assumed to facilitate grounding activities. Several key characteristics of collocated synchronous interactions have been identified by Olson and Olson (2000):
- rapid feedback: It allows for rapid corrections when there are misunderstandings or disagreements;

- multiple channels (visual, oral, etc.): it allows for several ways to convey complex message and provides redundancy;
- shared local context: a shared frame on the activities allows for mutual understanding about what is in other's mind;
- co-reference: gaze and gestures can easily identify the referent of deictic terms;
- spatiality of reference: both people and ideas (work objects) can be referred to spatially.

The importance of a shared context, in particular through mediating artefacts, has been put into evidence in co-design. Tang (1991) analysed the functions of collaborative drawing space in collocated design. The drawing space is an important resource for the group in mediating their interaction, e.g. moderating the turn taking, directing the group attention, providing additional context about who is contributing and what does it is mean. This function[4] reflects grounding via shared artefacts.

Distant design is mediated by various technologies which may affect grounding. Various media provide various kinds of cues which may imply various kinds and levels of effort for people to establish common ground. Clark and Brennan (1991) have identified several factors (called constraints) that can contribute to the construction of common ground via different media:

- Co-presence: A and B share the same physical environment.
- Visibility: A and B are visible to each other.
- Audibility: A and B communicate by speaking
- Cotemporality: B receives (messages) at roughly the same time as A produces.
- Simultaneity: A and B send and receive at once and simultaneously.
- Sequentiality: A's and B's turns cannot get out of sequence.
- Reviewability: B can review's A's messages.
- Revisability: A can revise messages for B before they are sent.

Collocated design meetings are characterised by co-presence, visibility, audibility, cotemporality, simultaneity and sequentiality. Co-presence ensures that co-designers have access to the same physical environment, and shared external representations, to support co-reference (in particular deictic ones) and shared context in grounding.

Studying how grounding and establishment of a shared context can take place in distant co-design, and also in asynchronous design, is an important direction of research which still remains to be explored. We will just here refer to a few studies with respect to different distant situations and technological supports.

In distant design, with video-conferencing (characterised by the same factors as collocated design meetings except co-presence) co-designers may be able to see each other without being able to see what each other is doing or looking at (functional visibility). It has been shown (Tang and Isaacs, 1993; Wittaker, 1995) that, in a complex task (such as design), face visibility is disturbing whereas functional visibility is helpful. Indeed functional visibility allows a local context to be shared and co-reference to be easier, thus facilitating grounding.

---

[4] However it should be noted that mediating artefacts have other functions (Schmidt and Wagner, 2002; Tang; 1991) such as storing information (artefact will later act as reminders of design principles, open problems, etc.; traces of activities; representation of design decisions) and expression of ideas (interactively creating representations of ideas in some tangible form, to enable the group to perceive, react to and build on them).

Détienne et al. (2004) analysed collaborative activities in co-design distant meetings with Netmeeting, shared CAD application and audio conferencing (characterised by audibility, relative contemporality and simultaneity, sequentiality). In this situation, there was a partial functional visibility of other co-designers work through the shared application. The authors found the same importance of grounding in this kind of situation as compared to collocated design meetings. However, extra interaction management activities were necessary to ensure explicit co-reference and shared context:
- information resource management: exchanges ensuring that team members are aware of the information under discussion, and have the same document version ( in their private or public spaces).
- screen management: exchanges ensuring that team members have a good visibility of the documents on the screen (like sketches).

Distant design, with email, chat and forum, (characterised by reviewability and revisability) is still a situation to be explored with respect to asynchronous co-design. In email and Chat there is no sequentiality as in face to face conversation where turns ordinarily form a sequence. A message and its reply may be separated by any number of irrelevant messages which makes grounding more difficult. We can analyse the quoting practice in exchanging messages (well identified in the computer-mediated communication field) as reflecting the reviewability factor as stressed by Clark and Brennan. This practice provides shared context in distant communication (Barcellini et al. 2005; Eklundh and Macdonald, 1994).

It should be noted that threaded chat may support the establishment of a shared local context by displaying messages in trees related to a topic in which messages and their replies are connected. However, the evaluation of this kind of environment has shown that the users complain that they loose control of what is going (Smith et al. 2000). They have more difficultly being conscious of what is happening in the various thematic conversations, thus loosing awareness of other people's actions.

*4.2. Perspectives clarification and convergence mechanisms*

In collaborative design, particularly co-design, designers apply various perspectives, linked to their domain of expertise but also to their responsibilities and interests, and usually negotiate in order to reach a solution which is "momentarily" agreed upon in the project. Most of the authors agree that a perspective or a viewpoint is strongly influenced by the domain area of the designer. Several participants see the design object differently according to the constraints specific to their discipline. In design, this is directly related to the Design Rationale (DR) approach.

Two mechanisms, tightly interrelated, are of particular interest in the co-construction of a negotiated solution:
- Perspectives clarification. It refers to the understanding of the reasoning behind design proposals and choices.  Of course, we could view this mechanism as a kind of grounding. In this paper, we restrict grounding to explanatory mechanisms (what proposal, how does it work) whereas perspective clarification refers to argumentation mechanisms (why this proposal).
- Mechanisms of convergence: the focus is on scenarios of interaction leading to convergence among co-designers.

*4.2.1 Perspectives clarification and the DR formalism*

Perspective clarification, by its focus on argumentation mechanisms, is tightly related to the DR approach. "A common assumption among early DR efforts using hypertext was that the DR formalism should serve as a common language for all of the different participants and perspective, in order to converge on shared understanding." (Conklin et al.2003). The DR approach in computer science have tempted to make explicit the reasoning behind design (Buckingham Shum et al. 1994; Conklin et al. 1991; Moran and Carroll, 1996).

Several DR notations have been developed to express the design reasoning as "arguments" about "issues". Among them, QOC and IBIS are probably the most well-known. The QOC notation (MacLean et al. 1991) distinguishes between Questions, Options and Criteria. The Design Space Analysis (DSA) approach that uses QOC consists in creating an explicit representation of a structured space of design alternatives and the considerations for choosing among them. It is a process of identifying key problems (Questions), and raising and justifying (via Criteria) design alternatives (Options). Criteria may be prescribed or derived, may be more or less discipline-specific, and may have various priority level attached according to the designers skills and interests and the problem at hand.

Buckingham et al.(1994) have reviewed the empirical evidence of the utility and usability of these techniques in design meetings. Whereas there is evidence that using an argumentative notation augments the reasoning of those who use it, there is also evidence that using such a notation impedes reasoning. The usability claim is not supported by the empirical evidence, which shows that using semiformal schemes for expressing knowledge introduces extra demands. As emphasised by Buckingham et al., "*the challenge for DR research is to find the most helpful, accessible representations of design reasoning both for initial developers and subsequent designers, which minimise the non-productive effort required to create them* [because capturing useful DR is bound to require some effort]."

Three main limitations for supporting perspectives clarification via the DR formalisms have been outlined.

The cost/benefit trade-off is important. Capturing DR requires some effort without immediate benefit for it. Thus designers tend to bypass DR clarification when it is possible. Two recent evolutions can improve this aspect.

One evolution consists in using DR as a management tool in the process-oriented approach of DR (Conklin and Burgess, 1991). In this way, it would be a productive effort with immediate feedback.

Another approach consists in the introduction of a mediator (person). The "facilitative approach" (Conklin et al. 2003) recommends that a skilled person (in the domain area), fluent with the formalisms and the hypertextools used for DR could act as a mediator between disciplines. However, it should be noted that the mediator is not only engaged in clarifying choices for decision making but also carry out a reflexive activity (Sauvagnac and Falzon, 2003).

Another limitation of the DR approach is that the identification of DR has been proven not to be easy in design meetings, especially in upstream design where the space of possible problems and solutions is still very open. Limiting the practice of DR to specific events such as downstream design meetings or technical review meetings meetings, in order to make them easier to accept (D'Astous et al. 2004) is one recent trend. In downstream design meetings and technical review meetings, previous tasks, for most of them individual design tasks, have already ensured partial structuring of the problem and solution spaces. Furthermore, argumentation plays an important role in these meetings and thus making DR explicit should be perceived as an intrinsic activity rather than as an added task.

Finally, DR was not integrated enough with the artefact under development. The artefact being "constructed" may be any product of the design process. A recent trend, called "construction driven argumentation" by Fischer et al. 1991, aims to improve such an integration.

- This integration can use an annotation technique or tags using metadata keyword (Conklin et al. 2003, Boujut and Laureillard, 2002), i.e., a secondary level of information attached to a main document, which has a special status in the project.
- In technical reviews, the to-be-revised artefact (D'Astous et al. 2004) will remain as documentation of the design project. DR methodology could facilitate the review of this artefact and could be used to extend the documentation, being integrated with the artefact itself. A design rationale could even replace the classic meeting minutes, which, most of the time, do not really reflect the actual nature of the discussions.

*4.2.2 Negotiation and convergence*

Another mechanism, the mechanism of convergence, ensures that the different disciplines will reach an agreement. It is also based on argumentation mechanisms. In the argumentative dialogue, a proposer expresses a proposal that will be argued about by presenting a certain amount of information substantiating the initial proposal. When everyone has a joint will to reach agreement, one talks about negotiation. Negotiation does not force a person to accept a solution, dialogue makes it possible to go towards one conclusion rather than another. For example, the conclusion can be a compromise between what each person wants.

Tools and methods supporting argumentation and, possibly, convergence are argumentation systems (for example, see Lonchamp 2000), voting mechanisms, or web user interface in which the document under discussion is tightly linked to a threaded discussion space (Sumner and Buckingham, 1988). These tools may use communicative acts or discourse ontology parameters (Uren et al. 2003): e.g., proves, refutes, is evidence for, is evidence again. They can be used to mediate face-to-face interaction, both physical and virtual (virtual meetings) and may be extended to asynchronous interactions.

Mediating activity through shared displays seems quite beneficial because it depersonalises conflict. Conklin et al. (2003, p 14) notice that: "When ideas and concerns are mediated via a shared display, challenges to positions assume a more neutral, less personal tone. It helps participants clarify the nature of their disagreement."

One important limitations of these approaches is that the generic model of negotiation implemented in these systems does not identify negotiation patterns with respect to particular team situation such as co-design meetings. Analysing the practices through which the convergence mechanisms occur could provide a model of design negotiation patterns. This is a direction of research rarely explored yet.

This has been undertaken in Détienne's work (Détienne et al. 2005; Martin et al. 2000b; Martin et al. 2001) who examined the negotiation patterns leading the participants to converge in multidisciplinary meetings in aeronautical design. They found a typical temporal negotiation pattern composed of three steps:

- Step1: analytical assessment mode of the current solution, i.e., systematic assessment according to constraints;
- Step 2: if step 1 has not led to a consensus, comparative or/and analogical assessment is involved. Comparative assessment mode consists in systematic comparison between alternative proposed solutions. Analogical assessment mode consists in transfer of knowledge acquired on a previous solution (accepted or not) in order to assess the current solution.
- Step 3: if step 2 has not led to a consensus, one (or several) argument(s) of authority[5] is (are) used.

They also found that three mechanisms were involved in sequences where converging among participants occurred: (1) explicitly negotiating constraints when the search for alternative solutions is in an impasse; (2) evoking shared knowledge concerning the evaluation of a source (previous solution developed for an analogous problem) or a previous alternative solution; (3) Argument of authority.

This kind of results may provide directions for the development of tools supporting negotiation. Distinguishing between negotiation phases and various assessment mechanisms could be a first step. However, further studies should extend this analysis in other design contexts.

## 5. Conclusion

This paper has presented an overview of current research issues on collaborative design by focusing on two great research challenges: managing task interdependencies and managing multiple perspectives. We have presented and discussed various approaches which cope with these challenges in collaborative design, their limitations and recent evolutions.

Research directions have been outlined with respect to collaboration in two design situations: coupling of work and organisation, informal communication and informal roles, and awareness in distributed design; establishment of common grounds, perspective clarification and convergence mechanisms in co-design.

Further studies could be conducted on these issues, adopting a generic approach of design. We believe that such an approach can be quite powerful, in the present state of research on collaborative design, in order to capitalise knowledge acquired in studies on collaborative design, which remain too often centred on results in a particular design domain with no reference to similar results in other design domains. However, ultimately, the approach will need to be refined to distinguish between types of design with respect to the nature of the design object (Visser, 2004).

---

[5] An argument can take the status of argument of authority depending on : the status, recognised in the organisation, of the discipline that expresses it; the expertise of the proposer; the "shared" nature of the knowledge to which it refers.